\documentclass[aps,pra,preprint,showpacs,floatfix,superscriptaddress]{revtex4}
\usepackage{graphicx} % Include figure files
\usepackage{amsmath}
\usepackage{bm}
\usepackage{dcolumn}% Align table columns on decimal point
\begin{document}

\title{Nuclear magnetic octupole moment and the hyperfine structure of
the $5D_{3/2,5/2}$ states of the Ba$^+$ ion}

\author{  K. Beloy}
\affiliation{Physics Department, University of Nevada, Reno, Nevada  89557}
\author{ A. Derevianko }
\affiliation{Physics Department, University of Nevada, Reno, Nevada  89557}
\affiliation{ School of Physics, University of New South Wales, Sydney 2052, Australia}
\author{ V. A. Dzuba }
\affiliation{ School of Physics, University of New South Wales, Sydney 2052, Australia}
\author{ G. T. Howell }
\affiliation{ Department of Physics, University of Washington, Seattle, Washington 98195}
\author{ B. B. Blinov }
\affiliation{ Department of Physics, University of Washington, Seattle, Washington 98195}
\author{ E. N. Fortson }
\affiliation{ Department of Physics, University of Washington, Seattle, Washington 98195}

\date{\today}
\begin{abstract}
The hyperfine structure of the long-lived $5D_{3/2}$ and $5D_{5/2}$ levels of Ba$^+$ ion is
analyzed. A procedure for extracting relatively unexplored nuclear magnetic moments $\Omega$ is presented.
The relevant electronic matrix elements are computed in the framework of the {\em ab initio} relativistic
many-body perturbation theory. Both the first- and the second-order
(in the hyperfine interaction) corrections to the energy levels are analyzed.
It is shown that a simultaneous measurement of the hyperfine structure of the entire
$5D_J$ fine-structure
manifold allows one to extract $\Omega$ without contamination from the second-order corrections.  Measurements to the required accuracy should be possible with a single trapped barium ion using sensitive techniques already demonstrated in Ba$^+$ experiments.
\end{abstract}

\maketitle

%\email{andrei@unr.edu}

%31.15.am Relativistic configuration interaction (CI) and many-body perturbation calculations
%AMO PACS:
%31.15.aj Relativistic corrections, spin-orbit effects, fine structure; hyperfine structure
%31.30.Gs Hyperfine interactions and isotope effects
%32.10.Fn Fine and hyperfine structure
%
%Nuclear Physics PACS:
%21.10.Ky Electromagnetic moments
%27.60.+j Properties of specific nuclei 90 <= A <= 149

\section{Introduction}

A nucleus as a source of the electro-magnetic fields is conventionally
described using a hierarchy of static electromagnetic
moments: magnetic dipole (M1,$\mu$), electric quadrupole (E2, $Q$),
magnetic octupole (M3,$\Omega$), etc.
Interaction of atomic electrons with these moments leads to
the hyperfine structure (HFS) of the atomic energy levels. While the first two
moments, $\mu$ and $Q$, have
been studied extensively, the octupole moments remain  relatively unexplored.

While octupole moments may be approximated using the nuclear-shell
model~\cite{Sch55}, the correct values depend strongly on nuclear
many-body effects and, in particular, on core-polarization mediated
by the nucleon spin-spin interaction~\cite{SenDmi02}.
\citet{SenDmi02} carried out a nuclear-structure calculation of
$\Omega$ for $^{209}$Bi. In this particular case, the polarization
effects enhance the shell-model values by a factor of three.
According to Dmitriev~\cite{DmiPrivate}, a systematic study of
octupole moments will help place constraints on the poorly-known
isoscalar part of nuclear spin-spin forces.  In another, even more
striking example, the deduced value of $\Omega$ of $^{133}$Cs is
forty times larger than the shell-model value~\cite{GerDerTan03}
(this has not been analyzed yet in nuclear theory). We note that in the case of
Cs, the shell-model value is strongly suppressed due to an
accidental cancellation between the orbital and spin contributions
of the valence proton to the magnetic-octupole moment.

Measuring the effects of the octupole moments on the hyperfine
structure was so far limited to a small number of atoms:
Cl~\cite{Sch57}, Ga~\cite{DalHol54}, Br~\cite{BroKin66},
In~\cite{EckKus57}, V~\cite{ChiPouGoo79V}, Eu~\cite{Chi91},
Lu~\cite{BreButRup85},  Hf~\cite{JinWakIna95}, and Bi~\cite{LanLur70}. Since deducing
$\Omega$ from a measurement requires knowing atomic-structure
couplings, previous analysis focused primarily on {\em isotopic
ratios} because the electronic coupling factor cancels out when
ratios of HFS constants are formed. An exception is the measurement
on the $6P_{3/2}$ state of $^{133}$Cs~\cite{GerDerTan03}, where
sufficiently accurate calculations are possible. In a recent
paper~\cite{BelDerJoh08}, we argued that an accurate deduction of
the octupole moments is feasible for metastable $^3P_2$ states of
alkaline-earth atoms.

Ba$^+$, being an atomic system with one valence electron outside a
closed-shell core, also presents a case where both high-accuracy
measurements and high-accuracy calculations are possible. The goal
of this paper is to analyze the hyperfine structure of the
$5D_{3/2}$ and $5D_{5/2}$ levels of Ba$^+$, and to show that Ba$^+$ is a
particularly favorable case for measuring octupole moments, for both theoretical
and experimental reasons.  Both $5D$ levels belong to
the same fine-structure manifold.
We take advantage of a simultaneous analysis of the hyperfine
structure of both levels and show that such an analysis allows
to eliminate the potentially troublesome second-order  hyperfine
electron structure term thus providing a powerful consistency test
for the measurements and calculations and possibly improving
the accuracy of deducting $\Omega$.
%We take advantage of a
%simultaneous analysis of the hyperfine structure of both levels
%% and show that such an analysis allows to eliminate certain
%to show that all potentially troublesome electronic-structure corrections
%(second-order hyperfine effects)can be canceled out,
%thus improving the accuracy of deducing $\Omega$.
Furthermore, $5D_{3/2}$ and $5D_{5/2}$ are each long lived metastable states
in Ba$^+$, with lifetimes of about $80\mathrm{s}$ and $30\mathrm{s}$ respectively.
Hyperfine intervals thus could be measured in principle to well below $0.1\mathrm{Hz}$,
much better than in previous octupole experiments; techniques to exploit the sensitivity
inherent in the 5D levels have already been developed using single trapped Ba$^+$~\cite{sherman05}.

Barium has two stable odd isotopes, 135 and 137. Both isotopes possess
nuclear ground states with spin of $I=3/2$ and of positive parity.
The values of the magnetic and quadrupole moments are
\begin{align}
^{135}\mathrm{Ba}:\, & \mu=0.837943 \, \mu_{N} \,,  Q=+0.160 \, \mathrm{b} \, ,\nonumber\\
^{137}\mathrm{Ba}:\, & \mu=0.937365 \, \mu_{N} \,,  Q=+0.245\, \mathrm{b} \, , \label{Eq:muQtabualted}
\end{align}
where $\mu_N$ is the nuclear magneton and b (barn) = $10^{-24}$
cm$^2$. Both isotopes have unpaired neutrons in the $d_{3/2}$
single-particle state, and from the single-particle (shell)
model~\cite{Sch55} we may estimate the octupole moment to be
\begin{equation}
\Omega^{\text{sp}}=0.164\mu_N\langle r^2 \rangle \approx 0.0385\, \mu_N\times\text{b} \, ,
\label{Eq:Omegasp}
\end{equation}
where we used the rms value of the nuclear radius $\langle r^2
\rangle^{1/2}=4.84 \, \mathrm{fm}$. Of course, the shell model is only an approximation.
For example, in contrast to the known properties~(\ref{Eq:muQtabualted})  for the two isotopes,
this model would produce identical values of the magnetic moments and
vanishing values of the quadrupole moments.

This paper is organized as follows.  First we recapitulate the theory
of the hyperfine structure of atomic levels, including octupole moments
and the second-order effects. We present specific formulae for the Ba case
and show how to extract the octupole constant $C$ from hyperfine intervals
(to be measured).  Further we compute the electronic structure factor required
for extracting the octupole moment from the constant $C$. Finally, we present a brief description of an experimental method with single Ba$^+$ to determine the octupole moments of $^{135}$Ba and $^{137}$Ba.  We follow the notation and formalism of the recent paper~\cite{BelDerJoh08}.
Unless specified otherwise, atomic units, $\hbar=|e|=m_e=1$, and
Gaussian electromagnetic units are employed throughout.

\section{Problem setup}
Hyperfine-interaction (HFI) Hamiltonian, describing coupling of electrons to various
nuclear moments may be represented in
the tensorial form of
\[
H_\text{HFI}=\sum_{k,q}\left(  -1\right)
^{q}T_{k,q}^{e}T_{k,-q}^{n}.
\]
Here the spherical tensors (of rank $k$) $T_{k,q}^{e}$ act on the
electronic coordinates. Tensor operators $T_{k,q}^{n}$  are the
components of the nuclear electric and magnetic 2$^{k}$-pole (MJ and
EJ; T,P-even) moment operators. In particular, the
conventionally-defined magnetic-dipole, electric-quadrupole, and
magnetic-octupole moments of the nucleus are proportional to the
expectation values of the zero-component ($q=0$) operators in the
nuclear stretched states $|I,M_I=I\rangle$: $ \mu = \langle
T_{k=1}^{n}\rangle_{I}$, $Q = 2\langle T_{k=2}^{n} \rangle_{I}$, and
$\Omega =-\langle T_{k=3}^{n} \rangle_{I}$. Explicit expressions for
the electronic operators and the corresponding reduced matrix
elements are tabulated in Ref.~\cite{BelDerJoh08}.

The conserved angular momentum $\mathbf{F}$ for the hyperfine
coupling  is composed from atomic, $\mathbf{J}$, and nuclear,
$\mathbf{I}$, angular momenta: $\mathbf{F}=\mathbf{I} +\mathbf{J}$.
It is convenient to work in a basis spanned by the eigenfunctions
$|\gamma IJFM_{F}\rangle$ which  is formed by coupling atomic,
$\left\vert \gamma JM_{J}\right\rangle$, and nuclear, $\left\vert
IM_{I}\right\rangle$, wave functions. Here  $\gamma$ encapsulates
remaining electronic quantum numbers. For $I=3/2$ each of the $5D_J$
levels splits into four hyperfine components: $5D_{3/2}$ has
hyperfine components $F=0,1,2,3$, and $5D_{5/2}$ has components
$F=1,2,3,4$.

Owing to the HFI's rotational invariance, a matrix element of the HFI
 in the $|\gamma IJFM_{F}\rangle$
basis is diagonal in the quantum numbers $F$ and $M_F$. If we limit
our system of levels to only the $5D_J$ fine-structure manifold, the
hyperfine components $F=1,2,3$ of the $5D_{3/2}$ and $5D_{5/2}$
levels become coupled.
 The intervals within each manifold may be parameterized using the
 conventional hyperfine constants $A$, $B$, $C$ and the second-order
corrections (in HFI) $\eta$ and $\zeta$.
 Constants $A$, $B$, $C$  are proportional to nuclear moments $\mu$, $Q$, and $\Omega$.
 M1-M1 correction $\eta$ is of the second order in $\mu$, and $\zeta$ comes from
 a cross-term between M1 and E2 parts of the HFI. Second-order
corrections are suppressed by a large energy denominator equal to
the fine-structure splitting between the $5D_J$ levels.

The energy intervals $\delta{W}^{(J)}_F=W^{(J)}_F-W^{(J)}_{F+1}$ within each fine-structure manifold
$5D_J$ are as follows.
For $5D_{3/2}$,
\begin{eqnarray}
\delta{W}^{(3/2)}_0&=&-A+B-56C
      +\frac{1}{100}\eta-\frac{1}{100}\sqrt{\frac{7}{3}}\zeta, \nonumber\\
\delta{W}^{(3/2)}_1&=&-2A+B+28C
      +\frac{1}{75}\eta, \nonumber\\
\delta{W}^{(3/2)}_2&=& -3A-B-8C
      +\frac{1}{300}\eta+\frac{1}{20}\sqrt{\frac{3}{7}}\zeta, \label{Eq:dW_3ov2}
\end{eqnarray}
and for $5D_{5/2}$,
\begin{eqnarray}
\delta{W}^{(5/2)}_1&=&-2A+\frac{4}{5}B-\frac{96}{5}C
      -\frac{1}{75}\eta, \nonumber\\
\delta{W}^{(5/2)}_2&=&-3A+\frac{9}{20}B+\frac{81}{5}C
      -\frac{1}{300}\eta-\frac{1}{20}\sqrt{\frac{3}{7}}\zeta, \nonumber\\
\delta{W}^{(5/2)}_3&=& -4A-\frac{4}{5}B-\frac{32}{5}C
      +\frac{2}{75}\eta+\frac{2}{25\sqrt{21}}\zeta. \label{Eq:dW_5ov2}
\end{eqnarray}
In the above equations the HFS constants $A$, $B$, and $C$ are all
specific to the state of consideration while $\eta$ and $\zeta$
represent the same second-order HFS constant.

If we assume all other second- and higher-order effects are
negligible (see justification in Section~\ref{Sec:discussion}), then we may solve for the HFS constants $A$, $B$, and
$C$ in terms of the HFS intervals and these two second-order
constants. Specifically, solving for the $C$ constants:
\begin{eqnarray}
C(5D_{3/2})&=&-\frac{1}{ 80}\delta{W}^{(3/2)}_0
              +\frac{1}{100}\delta{W}^{(3/2)}_1
              -\frac{1}{400}\delta{W}^{(3/2)}_2
              -\frac{1}{2000\sqrt{21}}\zeta,\nonumber\\
C(5D_{5/2})&=&-\frac{1}{ 40}\delta{W}^{(5/2)}_1
              +\frac{1}{ 35}\delta{W}^{(5/2)}_2
              -\frac{1}{112}\delta{W}^{(5/2)}_3
              +\frac{1}{200 \sqrt{21}}\zeta.
\label{Eq:C}
\end{eqnarray}
The $C$ constants do not depend on the M1-M1 $\eta$ correction,
as was proven in Ref.~\cite{BelDerJoh08} on general grounds.

It is possible to  use Eqs.~(\ref{Eq:C}) to cancel the constant
$\zeta$ and therefore eliminate the second-order effects from the
problem altogether. In doing so, we obtain the equation
\begin{eqnarray}
C(5D_{3/2})+\frac{1}{10}C(5D_{5/2})
           &=&-\frac{1}{  80}\delta{W}^{(3/2)}_0
              +\frac{1}{ 100}\delta{W}^{(3/2)}_1
              -\frac{1}{ 400}\delta{W}^{(3/2)}_2 \nonumber\\
            &&-\frac{1}{ 400}\delta{W}^{(5/2)}_1
              +\frac{1}{ 350}\delta{W}^{(5/2)}_2
              -\frac{1}{1120}\delta{W}^{(5/2)}_3 \, .
\label{Eq:CnoZ}
\end{eqnarray}
Since each of the constants $C$ is proportional to the same octupole
moment, knowing the hyperfine splitting inside each of the
fine-structure manifolds provides direct access to $\Omega$.

%------------------------------------------------------------------------------------
\section{Electronic-structure factors}

Provided that the measurements of the HFS intervals are carried out,
one could extract the octupole moment by computing the matrix
elements of the electronic coupling tensor $T_{3}^{e}$.
Specifically,
\[
C\left(  \gamma J\right)  =-\Omega~\left(
\begin{array}
[c]{ccc}%
J & 3 & J\\
-J & 0 & J
\end{array}
\right)  ~\langle\gamma J||T_{3}^{e}||\gamma J\rangle \, .
\]
We may also compute the second-order M1-E2 HFS correction $\zeta$ by
computing the off-diagonal matrix elements of the electronic
coupling tensors $T_{1}^{e}$ and $T_{2}^{e}$. For $I=3/2$, $\zeta$
is given by
\begin{equation*}
\zeta=\frac{20}{\sqrt{3}}\frac{\mu Q
\langle{5D_{3/2}||T^e_1||5D_{5/2}}\rangle
\langle{5D_{5/2}||T^e_2||5D_{3/2}}\rangle}{E_{5D_{5/2}}-E_{5D_{3/2}}}.
\end{equation*}
Matrix elements of the electronic tensors are given in Ref.~\cite{BelDerJoh08}.

% The numerical calculations were carried out using a dual kinetic-balance (DKB) basis
% set~\cite{ShaTupYer04}, as described in Ref.~\cite{BelDer08}.

To calculate the electronic-structure factors for the HFS constants
$C(5D_{3/2})$ and $C(5D_{5/2})$ and for the second-order term
$\zeta$ we employ the correlation potential
method~\cite{DzuFlaSil87} using all-order correlation correction
operator $\hat \Sigma^{(\infty)}$ as suggested in
Refs.~\cite{DzuFlaSus89,DzuFlaKra89}. The method was used for Ba$^+$
previously~\cite{DzuFlaGin01} for accurate calculation of the parity
non-conservation. It is also known that the method produces accurate
results for the magnetic dipole hyperfine structure constants of
alkali atoms (see, e.g. Ref.~\cite{DzuFla00}).

Calculations start from the Hartree-Fock procedure for the closed-shell
Ba$^{2+}$ ion
\begin{equation}
  (\hat H_0 - \epsilon_c)\psi_c = 0,
\label{RHF}
\end{equation}
where $\hat H_0$ is the single-electron relativistic Hartree-Fock
(RHF) Hamiltonian
\begin{equation}
  \hat H_0 = c{\bf \alpha \cdot \hat p} +(\beta-1)mc^2 - Ze^2/r + \hat V_{core},
\label{HHF}
\end{equation}
index $c$ in Eq.(\ref{RHF}) numerates core states, and $\hat V_{core}$ is the sum of
the direct and exchange self-consistent potential created by $Z-2$
core electrons.

States of the external electron are calculated using the equation
\begin{equation}
  (\hat H_0 +\hat \Sigma - \epsilon_v)\psi_v = 0,
\label{Brueck}
\end{equation}
which differ from the equation for the core (\ref{RHF}) by an
extra operator $\hat \Sigma$.
% which is the correlation correction
%operator or the correlation potential.
The so-called correlation operator
$\hat \Sigma$ is defined in such a way that in the lowest order
the correlation correction to the energy of the external
electron is given as an expectation value of the $\hat \Sigma$
operator
\begin{equation}
  \delta \epsilon_v = \langle \psi_v | \hat \Sigma | \psi_v \rangle.
\label{Sigma}
\end{equation}
The correlation potential $\hat \Sigma$ is a non-local operator
which is treated in the Hartree-Fock-like equations (\ref{Brueck})
the same way as a non-local exchange potential. Solving these
equations we get the energies and the orbitals which include
correlations. These orbitals are usually called {\em Brueckner
orbitals}.

We use the Feynman diagram technique~\cite{DzuFlaSus89} and B-spline
basis set~\cite{JohBluSap88} to calculate $\hat \Sigma$. The
many-body perturbation theory (MBPT) expansion for $\hat \Sigma$
starts from the second-order and has the corresponding notation
$\hat \Sigma^{(2)}$. However, we also include two dominating classes
of the higher-order diagrams into the calculation of $\hat \Sigma$,
as described in Ref.~\cite{DzuFlaSus89}. These higher-order effects
are: (1) screening of Coulomb interaction between core and valence
electrons by other core electrons and (2) an interaction between an
electron excited from atomic core and the hole in the core created
by this excitation. Both these effects are included in all orders
and corresponding $\hat \Sigma$ is called $\hat \Sigma^{(\infty)}$.
Another class of higher-order correlations is included
in all orders when the equations (\ref{Brueck}) are iterated
for the valence states. These higher-order effects are
proportional to $\langle \hat \Sigma \rangle^2$,
$\langle \hat \Sigma \rangle^3$, etc.

The effect of the second- and higher-order correlations on the
energies of Ba$^+$ is illustrated by the data in Table ~\ref{tb:en}.
As one can see, the inclusion of the correlations leads to
systematic improvement in the accuracy for the energies. Note that
since solving the equations (\ref{Brueck}) for Brueckner orbitals
produces not only the energies but also the wave functions of the
external electron, the better accuracy for the energy should
translate into better accuracy for the wave function and for the
matrix elements. Therefore, we can try to improve the wave function
even further by fitting the energies to the experimental values by
rescaling  the $\hat \Sigma$ operator in Eq.~(\ref{Brueck}). This is
done by replacing $\hat \Sigma$ by $f\hat \Sigma$, where rescaling
parameter $f$ is chosen to fit experimental energies. The values of
$f$ for different states of Ba$^+$ are listed in Table~\ref{tb:en}.

\begin{table}[h]
\caption{Removal energies of Ba$^+$ in different approximations [cm$^{-1}$].}
\label{tb:en}
\begin{tabular}{ccccccc}
State & $J$ & RHF & $\Sigma^{(2)}$ &  $\Sigma^{(\infty)}$ & fitted\tablenotemark[1]
& Expt.\tablenotemark[2] \\
\hline
$6S$ & 1/2  &  75339 &   82227 &   80812 &   80685 &   80687 \\
$6P$ & 1/2  &  57265 &   61129 &   60584 &   60442 &   60425 \\
$6P$ & 3/2  &  55873 &   59351 &   58863 &   58735 &   58734 \\
$5D$ & 3/2  &  68138 &   77123 &   76380 &   75816 &   75813 \\
$5D$ & 5/2  &  67664 &   76186 &   75543 &   75004 &   75012 \\
\hline
\end{tabular}
\tablenotetext[1]{$f(6s)$=0.978, $f(6p)$=0.960, $f(5d)$=0.934}
\tablenotetext[2]{NIST, Ref.~\cite{NIST_ASD}}
\end{table}

To calculate the HFS constants we need to include extra fields
which are the fields of the nuclear P-even electromagnetic moments
such as magnetic dipole, electric quadrupole, etc.
This is done in the self-consistent way similar to the RHF
calculations for the energies in the frameworks of the well-known
random-phase approximation (RPA).
Corresponding equations have the form
\begin{equation}
  (\hat H_0 - \epsilon_c)\delta \psi_c = -(\hat F + \delta \hat V_{core}) \psi_c,
\label{RPA}
\end{equation}
where $\hat F$ is the operator of external field, $\delta \psi_c$ is
the correction to the core state due to the effect of external field
and $\delta \hat V_{core}$ is the correction to the self-consistent
Hartree-Fock potential due to the change in field-perturbed core
states. The RPA equations (\ref{RPA}) can be considered as a
linearized (in external field) expansion of the RHF equations
(\ref{RHF}); these are also solved self-consistently for all the
core states. This corresponds to the inclusion of the so-called {\em
core polarization} (CP) effect. Matrix elements for states of the
external electron are given by
\begin{equation}
  \langle \psi_v | \hat F + \delta \hat V_{core} | \psi_v \rangle.
\label{me}
\end{equation}
Dominant correlations are included by simply using the Brueckner
orbitals as the wave functions $\psi_v$ in (\ref{me}).
There are, however, correlation corrections to the matrix
elements which are not included into (\ref{me}). These are the
{\em structure radiation} (SR) and the effect of normalization of
the many-electron wave function~\cite{DzuFlaSil87}. Structure radiation
can be described as a contribution due to the change in
$\hat \Sigma$ caused by the effect of the external field:
\begin{equation}
  \langle \psi_v | \delta \hat \Sigma | \psi_v \rangle.
\label{SR}
\end{equation}
We calculate SR and renormalization contributions using the MBPT
similar to the third-order calculations presented in
Ref.~\cite{JohLiuSap96} (second-order in Coulomb interaction and
first-order in external field). However, we use the ``dressed''
operators of the external field: $\hat F + \delta \hat V_{core}$
rather than just $\hat F$ as in Ref.~\cite{JohLiuSap96}. Therefore,
core polarization effect is included in all orders in the SR and
renormalization calculations. We also use two different basis sets
of single-electron states. One is the dual kinetic-balance basis
(DKB) set~\cite{ShaTupYer04,BelDer08}, and another is the B-spline
basis set developed at the University of Notre
Dame~\cite{JohBluSap88}.

The results of the calculations are presented in
Table~\ref{tb:Chfs}. Here the RHF approximation corresponds to the
$\langle \psi_v^{\text{HF}} | \hat F | \psi_v^{\text{HF}} \rangle$
matrix elements with the Hartree-Fock orbitals $\psi_v^{\text{HF}}$.
RPA approximation corresponds to the $\langle \psi_v^{\text{HF}} |
\hat F + \delta \hat V_{core}| \psi_v^{\text{HF}} \rangle$ matrix
elements. Brueckner + CP approximation corresponds to the $\langle
\psi_v^{\text{Br}} | \hat F + \delta \hat V_{core}|
\psi_v^{\text{Br}} \rangle$ matrix elements with Brueckner orbitals
$\psi_v^{\text{Br}}$, etc. The values of  the SR and renormalization corrections
listed in Table~\ref{tb:Chfs} were computed with the DKB basis set.

\begin{table}[h]
\caption{Magnetic octupole hyperfine structure constant $C$ of the
$5D_{3/2}$ and $5D_{5/2}$ states of Ba$^+$ and off-diagonal matrix
elements of the magnetic dipole and electric quadrupole operators in
different approximations.} \label{tb:Chfs}
\begin{tabular}{ldddd}%{l r r r r}
Approximation & \multicolumn{1}{c}{$C(5D_{3/2})$} &
\multicolumn{1}{c}{$C(5D_{5/2})$} & \multicolumn{1}{c}{$\langle
5D_{3/2}||T^{(e)}_1|| 5D_{5/2} \rangle$}
& \multicolumn{1}{c}{$\langle 5D_{5/2}||T^{(e)}_2|| 5D_{3/2} \rangle$} \\
& \multicolumn{1}{c}{$\rm{kHz}/\left(\Omega/(\mu_N \times
\rm{b})\right)$} & \multicolumn{1}{c}{$\rm{kHz}/\left(\Omega/(\mu_N
\times \rm{b})\right)$} & \multicolumn{1}{c}{MHz/$\mu_N$}
& \multicolumn{1}{c}{MHz/b} \\
\hline
RHF                                       & -0.4294 & -0.1514 &   -95 & 180 \\
RPA $\equiv$ RHF + CP & -0.5843 &  0.9636 & -1360 & 184 \\
$\Sigma^{(2)}$ + CP            & -0.6863 &  0.9254 & -1496 & 222 \\
$\Sigma^{(\infty)}$ + CP       & -0.6822 &  0.9244 & -1489 & 220 \\
Energy fitting (Br)     & -0.6758 &  0.9282 & -1481 & 218 \\
SR                  &  0.0842 & -0.8472 &   280 &  14 \\
Norm                 &  0.0178 & -0.0287 &    42 &  -5 \\
Total  (Br+SR+Norm)  & -0.5738 &  0.0523 & -1160 & 227 \\
\hline
\end{tabular}
\end{table}

\section{Extracting octupole moment}
\label{Sec:discussion}
Our final results for the magnetic octupole hyperfine structure constants are
\begin{eqnarray}
  C(5D_{3/2}) &=& -0.585(11)
   \left( \frac{\Omega}{\mu_N \times \mathrm{b}} \right) \,  {\rm kHz} \, , \nonumber\\
  C(5D_{5/2}) &=&  0.036(16)
  \left( \frac{\Omega}{\mu_N \times \mathrm{b}} \right) \,  {\rm kHz} \,.
\label{Chfs}
\end{eqnarray}
Here central values and the errors are found from the scattering of the
results due to effects of energy fitting and change of basis for the SR
and renormalization calculations. Notice that the error bars are purely
theoretical and reflect the fact that only certain classes of diagrams
are included in the calculations. In particular, there are strong cancelations
between various contributions to the  $C(5D_{5/2})$ constant, leading to
a large, 45\%, uncertainty in the value of this constant.

The above error estimates are consistent with the general trend for
the experimentally known constants $A$ and $B$ of the $5D_J$
states~\cite{SilBorDeB86}. Our employed method is off by as much as
10\% for $A(5D_{3/2})$ and 30\% for $A(5D_{5/2})$. The computed
values of $B$ generally agree at the level of a few per cent with
the experiment. Another insight comes from understanding that the
theoretical method includes the RPA and Brueckner chains to ``all
orders''; as long as these classes of diagrams dominate, the
theoretical accuracy is excellent. By contrast, in case of the
$C(5D_{5/2})$ the result is accumulated due to remaining SR and Norm
diagrams (see Table~\ref{tb:Chfs}), which are computed nominally in
the third-order MBPT only. This explains the relatively poor
accuracy for the $5D_{5/2}$ states. As demonstrated in Ref.~\cite{Sah06} (at least for the
 constants $A$ and $B$) the theoretical accuracy could be improved to 1\%  by employing the relativistic coupled-cluster method.

The results, Eqs.~(\ref{Chfs}), may be used to extract the values of
the nuclear magnetic octupole moment from the measurements. For
example, if equation~(\ref{Eq:CnoZ}) is used then
\begin{equation}
  C(5D_{3/2})+\frac{1}{10}C(5D_{5/2}) = -0.581(13) \,
  \left( \frac{\Omega}{\mu_N \times \mathrm{b}} \right) \,  {\rm kHz} .
\label{C5d}
\end{equation}
Alternatively, one can use the first equation of (\ref{Eq:C}). Then
the correction due to the second-order term $\zeta$ needs to be
taken into account. With the values of the magnetic dipole and
electric quadrupole HFS matrix elements presented in the last
columns of Table~\ref{tb:Chfs} this correction reads
\begin{equation}
  \Delta C(5D_{3/2}) = -\frac{1}{2000\sqrt{21}}\zeta =
  \left\{ \begin{array}{lll} 1.84 & {\rm Hz} & ^{135}{\rm Ba}^+ \\
                            3.17 & {\rm Hz} & ^{137}{\rm Ba}^+ \\
\end{array}
\right.
\label{Eq:final}
\end{equation}
Notice that in this second scenario we advocate using the $5D_{3/2}$
hyperfine manifold for extracting the nuclear octupole moment
because of the poor theoretical accuracy of computing electronic
couplings for the $5D_{5/2}$  state.

We may evaluate the relative influence of $\Omega$ on the HFS by
using the single-particle (shell-model) estimate for the nuclear
octupole moment, Eq.~(\ref{Eq:Omegasp}); we arrive at
\begin{eqnarray*}
C(5D_{3/2})^\mathrm{s.p.} &\approx& -23 \, \rm{Hz} \, , \nonumber\\
C(5D_{5/2})^\mathrm{s.p.} &\approx& 1.4 \, \rm{Hz} \, .
\end{eqnarray*}
We see that the second-order correction, $\Delta C(5D_{3/2})$, is
below the anticipated value of the constant.

At this point we briefly discuss the effect of all other second- and higher-order terms past $\eta$ and $\zeta$, which until this point have been assumed negligible. One might expect other second-order dipole-dipole terms which mix in states outside of the $5D$ fine structure manifold to have an appreciable effect on the hyperfine structure. However, the proof in Ref.~\cite{BelDerJoh08} can easily be generalized to show that {\em no} second-order dipole-dipole terms enter into the equations for the $C$ constants, Eqs.~(\ref{Eq:C}). Furthermore, it is found that the leading third-order term, the dipole-dipole-dipole term mixing the fine structure levels, drops out of Eq.~(\ref{Eq:CnoZ}) along with $\zeta$. Therefore, we can expect the largest terms neglected from Eq.~(\ref{Eq:CnoZ}) to be the second-order dipole-octupole and quadrupole-quadrupole terms; we have estimated these effects to both be at the $\sim 10^{-3}$ Hz level. 
This provides sufficient confirmation that all second- and higher-order 
terms may be neglected in our proposed scheme of extracting
 the octupole moment.
%In addition, the correction to the second-order terms due to the other 
%electronic intermediate states (e.g., $6D_J$) are suppressed by the ratio 
%of the fine-structure splitting in the $5D$ manifold to a much larger 
%energy separation between the $5D$ and $6D$ states. 
%Such corrections are expected to enter at the level below a few per 
%cent to the computed values of the second-order corrections, 
%and are below the accuracy of the present calculation.

Finally, using the single-particle approximation for the nuclear
octupole moment, we obtain an estimated value for the left hand side
of Eq.~(\ref{Eq:CnoZ}) of $-22$ Hz.
Assuming a conservative value of 1 Hz uncertainty in the $5D$ HFS intervals yields an overall uncertainty of $0.017$ Hz in the right hand side of
Eq.~(\ref{Eq:CnoZ}).  In the next section we describe how such measurements should be capable of much smaller uncertainties, 0.1 Hz or  better.  Thus we may conclude
that HFS interval measurements with readily attainable accuracy,
combined with the theoretical result (\ref{C5d}), would be capable
of extracting an octupole moment of the estimated size for the
$^{135}\text{Ba}$ and $^{137}\text{Ba}$ nuclei.

\section{Experimental Possibilities}

The measurements can be carried out by a technique similar to one already used to study transitions among sublevels of the $5D_{3/2}$ state of Ba$^+$~\cite{sherman05}, in which optical pumping is used to place the ion in a particular sublevel, and an RF transition to another sublevel is detected by the effect of ``shelving" as described below.   Such measurements are performed on a single ion held by radio frequency electric fields in a 3-dimensional effective potential well typically $\approx 100 \mathrm{eV}$ deep, with the ion at the bottom of the well after being laser cooled to a temperature $\cong 10^{-3} \deg \mathrm{K}$, with an orbital diameter $\cong 10^{-2}\mu$m.

\begin{figure}
\includegraphics[scale=0.60]{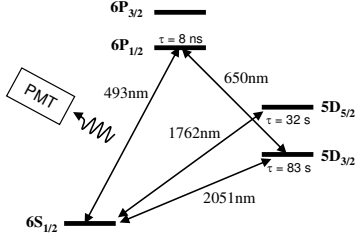}
\caption{\label{fig:levels} The lowest $S$, $P$, and $D$ states
of $\mathrm{Ba}^+$, showing the cooling 493 nm and cleanup 650 nm transitions plus the 2051 nm and 1762 nm E2 transitions  to the metastable $D$ states. }
\end{figure}

The electronic energies of the lowest $S$, $P$, and $D$ states
of Ba$^+$ are shown in Fig.~\ref{fig:levels}.  The cooling laser operates on the
$6S_{1/ 2} - 6P_{1/ 2}$ allowed $E1$ absorption line
near 493 nm, mistuned slightly to the red of resonance to
effect Doppler cooling.  A `cleanup' laser beam operates
at the $6P_{1/ 2} - 5D_{3/ 2}$ transition near 650 nm
to keep the ion from getting stuck in the metastable $D$ state and lost to the cooling process.

\begin{figure}
\includegraphics[scale=0.60]{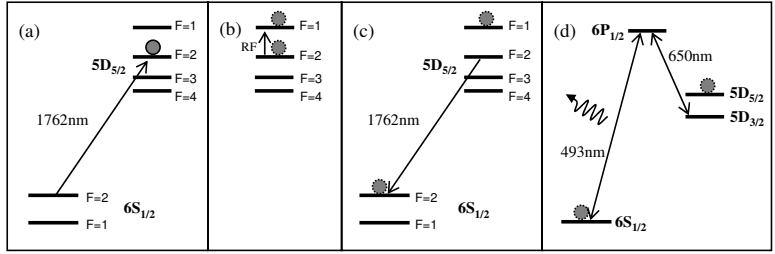}
\caption{\label{fig:shelve} Procedure for measuring the hyperfine intervals in the $5D_{5/2}$ state. (a) First, a 1762 nm resonant laser pulse transfers the ion into the $F=2$ hyperfine sublevel of the $5D_{5/2}$ state. (b) The RF is then applied to drive a hyperfine transition in the $5D_{5/2}$ state. (c) The second pulse of the 1762 nm laser depopulates the $5D_{5/2}$ $F=2$ level. (d) To determine if the hyperfine transition in the $5D_{5/2}$ state occurred, the cooling and the cleanup lasers are turned on and any ion fluorescence is detected.  Absence of fluorescence indicates that the RF is on resonance and caused the transition to take place.
 }
\end{figure}

To measure the hyperfine splitting of either $5D$ state, the ion can be initially placed in the ($F_g =2$, $M_F = 0$) Zeeman sublevel of the $6S_{1/ 2}$ ground state by optical pumping with a polarized 493 nm beam.  As shown in Fig.~\ref{fig:shelve}, in the case of the $5D_{5/ 2}$ measurement the ion is then transferred to a particular hyperfine sublevel ($F$, $M_F$) of $5D_{5/ 2}$ by applying a pulse of resonant 1762 nm light. RF field coils are then turned on for driving a $\Delta F = \pm 1$ transition, after which a second pulse of the 1762 nm laser will transfer from ($F$, $M_F$) back to the ground state.  The ion therefore ends up in the ground state if there was no hyperfine transition, and in $5D_{5/ 2}$ if the RF is on resonance and the hyperfine transition was successful.  In the former case there will be fluorescence when the ion is illuminated by the 493 nm/650 nm lasers while in the latter case there will not be fluorescence; the ion is `shelved' in the $5D_{5/ 2}$ state.  The process is repeated for a range of RF frequencies and a hyperfine transition resonance curve is acquired.  For the $5D_{3/ 2}$ measurement, the same procedure is followed with 2051 nm resonant light to populate $5D_{3/ 2}$ sublevels, but an extra step is needed at the end -- the shelving of the $6S_{1/ 2}$ state population to the $5D_{5/ 2}$ state.

\begin{table}[h]
\caption{Zeeman splitting coefficients defined in Eq.~(\ref{Eq:zeeconst}). }
\label{tb:zee}
\begin{tabular}{cccc}
$M_F\:\:$ & $a_1 \:($MHz/G$)\:\:$ & $a_2 \:($MHz/G$)\:\:$ & $ a_3 \:($MHz$^2$/G$^2)\:\:$  \\
\hline
$3$ & 3.359  &  0.2099 &   0.7052  \\
$2$ & 2.239  &  0.1400 &   1.153  \\
$1$ & 1.120  &  0.06998 &   1.422  \\
$0$ & 0      &  0       &   1.511   \\
$-1$ & -1.120 &  -0.06998 &   1.422 \\
$-2$ & -2.239 &  -0.1400 &   1.153  \\
$-3$ & -3.359 &  -0.2099 &   0.7052  \\
\hline
\end{tabular}
\end{table}

In the previous measurements of Zeeman transitions among $5D_{3/ 2}$ sublevels, sensitivities of a few Hz were achieved~\cite{sherman05}, limited by incompletely shielded magnetic field fluctuations.  In the hyperfine measurements proposed here, this source of broadening can be eliminated by using transitions that have a weak magnetic field dependence, such as $M_F = 0 \rightarrow M_F = 0$.  In practice, it is often easiest to perform laser cooling in nonzero magnetic fields (typically $B \cong 1$G) to avoid formation of inefficiently cooled dark states.  Such modest magnetic fields will introduce only a small $B^2$ dependence in the $0 \rightarrow 0$ transition when the energy separations between hyperfine levels are sufficiently large.  In fact the Zeeman effect for $B = 1$G can indeed be considered a small perturbation to the hyperfine splitting for all the $5D_{3/ 2}$  states and for the $F =1$ and $F = 2$ states of the $5D_{5/ 2}$  manifold.

\begin{figure}[tbp] % float placement: (h)ere, page (t)op, page (b)ottom, other (p)age
  \centering
  % file name: C:/Documents and Settings/students/Desktop/Zeeman_diagram_3.pdf
  \includegraphics[bb=85 176 674 473,width=5.94in,height=4in,keepaspectratio]{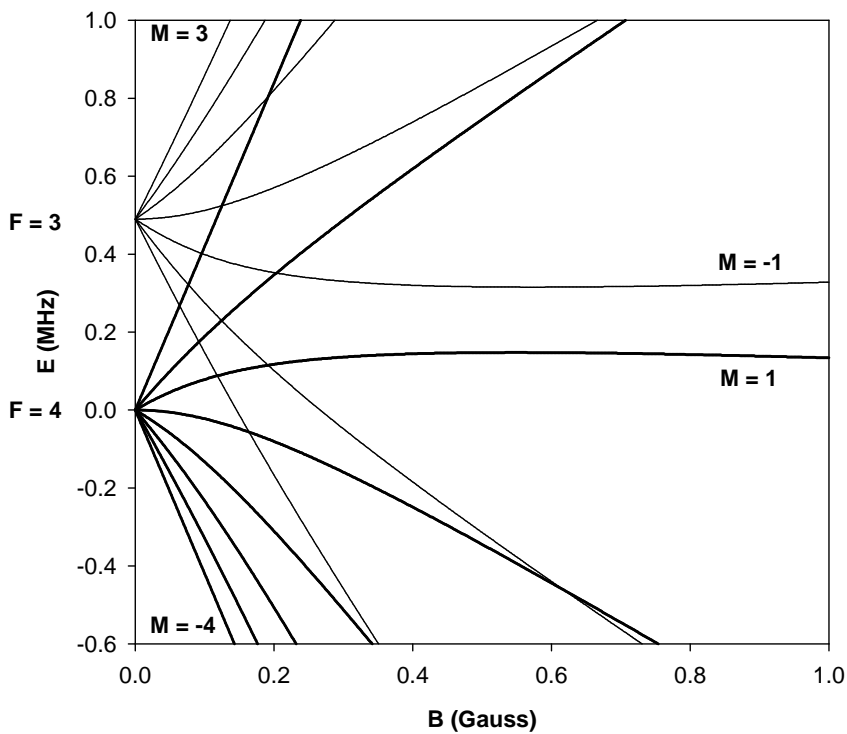}
  \caption{$D_{5/2}\;F=3$ and $F=4$ hyperfine Zeeman levels.  Energies are measured relative to the $F=4$ energy at zero magnetic field.  Very small mixing of the $F=3$ state with the $F=2$ state has been neglected.}
  \label{fig:Zeeman}
\end{figure}

However, for the $5D_{5/ 2}$   $F=3$ and $F=4$ states the situation is more complicated because the hyperfine splitting between these states is small, $\cong 0.49$ MHz~\cite{SilBorDeB86}, so the Zeeman Hamiltonian cannot be considered a weak perturbation to the hyperfine Hamiltonian.  The Zeeman and hyperfine Hamiltonians must therefore be treated on an equal footing.  The manifold of $F=3$ and $F=4$ states are then all degenerate in zeroth order, and the first-order energy shifts due to  $H_{Zeeman}+ H_{hyp}$ are obtained by diagonalizing the matrix of  $H_{Zeeman}+ H_{hyp}$  within the manifold of $F=3$ and $F=4$. In the $|IJFM_{F}\rangle$ basis, the matrix is 2x2 block diagonal, with each block having a given value of $M_F$.  The eigenvalues are then given by
\begin{equation}
  E_\pm=\frac{1}{2}E_{3}+a_1B\pm\sqrt{\frac{1}{4}E_{3}^{2} + a_2E_3B + a_3B^{2}}
\label{zee}
\end{equation}
where $E_+$ reduces to the $F=3$ hyperfine energy $E_3$ for $B=0$ and $E_-$ reduces to the $F=4$ hyperfine energy, which is here taken as zero.  The values of $a_1$, $a_2$, and $a_3$ are given in Table III.  These are related to the matrix elements of the Zeeman Hamiltonian $H'$ as follows
\begin{eqnarray}
a_1B=\frac{1}{2}\left(\:\langle4,M_F| H'|4,M_F\rangle+\langle
	3,M_F| H'|3,M_F\rangle\:\right),\nonumber\\
a_2B=-\frac{1}{2}\left(\:\langle4,M_F| H'|4,M_F\rangle-\langle
	3,M_F| H'|3,M_F\rangle\:\right),\nonumber\\
a_3B^2=\frac{1}{4}\left(\:\langle4,M_F| H'|4,M_F\rangle	-\langle
	3,M_F| H'|3,M_F\rangle\:\right)^2+\left(\:\langle
	4,M_F| H'|3,M_F\rangle\:\right)^2.
\label{Eq:zeeconst}
\end{eqnarray}
A portion of the graph of the energy levels as a function of $B$ is given in Fig.~\ref{fig:Zeeman}.  Note that for even small magnetic fields there is a great deal of mixing between the $F=3$ and $F=4$ states, and there are two energy levels, originating from $F=3,\:M_F=-1$ and $F=4,\:M_F=1$, that are almost field-independent for $B$ in the range of $\simeq0.6 G$ -- $1.2 G$.  Each of these levels can be connected to the $F=2$, $M_F = 0$ level by an RF transition, with very weak dependence on $B$.  Likewise, the desired zero-field hyperfine intervals can be extracted from these RF measurements using only a relatively low resolution determination of $B$ by a field-dependent Zeeman resonance.  Thus measurement of all hyperfine intervals to 0.1 Hz seems feasible.

We have shown that a simultaneous measurement of the hyperfine splittings in the 5D$_{3/2}$ and the 5D$_{5/2}$ fine
structure levels of Ba$^{+}$ allows one to unambiguously extract the
value for the nuclear magnetic octupole moment. We performed the \textit{ab initio} calculations of the relevant
matrix elements in the framework of relativistic many-body perturbation theory, analyzing the first- and the second-order
corrections to the hyperfine energy levels. We have also outlined an experimental procedure for measuring
the hyperfine intervals to the required accuracy with single trapped Ba$^{+}$ ions.

\acknowledgements
We would like to thank V. F. Dmitriev, V. Flambaum,  and Jeff Sherman for discussions.
AD was supported in part by the US Dept.~of State Fulbright fellowship to Australia
and would like thank the School of Physics of the University of New South Wales for hospitality.
The work of KB and AD was supported in part by National Science Foundation
grant No. PHY-06-53392 and that of ENF by National Science Foundation
grant No. PHY-04-57320.

%\bibliography{all,addbib}

\end{document}